\documentclass[11pt,a4paper]{article}

\usepackage[a4paper,margin=2.2cm]{geometry}
\usepackage[T1]{fontenc}
\usepackage[utf8]{inputenc}
\usepackage{lmodern}
\usepackage{microtype}

\usepackage{amsmath}
\usepackage{amssymb}

\usepackage{graphicx}
\usepackage{overpic}
\usepackage{subcaption}
\usepackage{float}
\usepackage[section]{placeins}
\usepackage{flafter}

\graphicspath{{figs/}}

\usepackage{algorithm}
\usepackage{algpseudocode}

\usepackage[hidelinks]{hyperref}

\title{
Local Variance-Based Calibration of Programmable Photonic
Interferometer Meshes
}

\author{
G{\"o}khan Elmas$^{1,*}$,
Igor A. Litvin$^{1}$,
and Janis N{\"o}tzel$^{1,\dagger}$
\\[0.6em]
\small
$^{1}$Emmy-Noether Group Theoretical Quantum Systems Design,
Technical University of Munich, Munich, Germany
\\[0.3em]
\small
$^{*}$\href{mailto:gokhan.elmas@tum.de}{gokhan.elmas@tum.de}
\qquad
$^{\dagger}$\href{mailto:janis.noetzel@tum.de}{janis.noetzel@tum.de}
}

\date{}

\begin{document}

\maketitle

\begin{abstract}
Programmable photonic interferometer meshes are powerful platforms for
implementing reconfigurable linear optical transformations, but their performance
depends critically on accurate calibration of Mach--Zehnder interferometers and
phase shifters. Conventional calibration methods often require node isolation,
dedicated routing paths, orthogonal training states, reference channels, or prior
phase--voltage characterization, all of which become increasingly difficult in
large thermally tuned meshes.

Here we introduce a local variance-based self-calibration method for programmable
photonic interferometer meshes using intensity-only measurements. The method
applies controlled phase perturbations and identifies calibration points from
minima of the measured output-power variance. For Mach--Zehnder interferometers,
the variance follows a characteristic $|\sin\theta|$ dependence, allowing bar and
cross operating points to be identified without relying on conventional node
isolation. For phase shifters, operation under balanced interference yields a
complementary $|\cos\phi|$ variance signature, enabling quadrature calibration
from the same statistical principle.

We experimentally validate the method on an $8\times8$ silicon-nitride
programmable photonic processor using a fully automated two-stage procedure.
Starting from random initial phase settings, all Mach--Zehnder interferometers
are calibrated first, followed by phase-shifter calibration under
balanced-interference conditions. As a system-level test, we implement an
embedded $4\times4$ Hadamard transformation on the $8\times8$ processor using a
Clements decomposition. Compared with the manufacturer-provided factory
calibration, the variance-based calibration improves output uniformity and
temporal stability across all channels.

These results establish local output variance as a simple calibration observable
for programmable photonic meshes. The approach is intensity-only, compatible
with discrete random phase ensembles, and does not require conventional node
isolation or orthogonal training fields, making it a practical calibration
primitive for scalable self-stabilizing photonic processors.
\end{abstract}

\vspace{0.5cm}

\section{Introduction}

Programmable photonic interferometer meshes composed of Mach--Zehnder
interferometers (MZIs) and phase shifters provide a universal platform for
implementing arbitrary linear optical transformations
~\cite{Miller2013,Reck1994,Clements2016,Carolan2015}. These systems have enabled
a broad range of applications in optical computing, neuromorphic processing,
quantum information processing, microwave photonics, and coherent optical
signal processing
~\cite{Shen2017,Hamerly2019,Wang2020,Perez2017,Litvin2025JLT,
Litvin2026Joint}. More broadly, programmable photonic circuits have emerged as
a general-purpose hardware paradigm in which meshes of tunable couplers and
phase shifters can be software-configured to implement different optical
functions~\cite{Bogaerts2020}. Experimental demonstrations of universal linear
circuits and large-scale quantum photonic processors have further established
interferometric meshes as a practical architecture for reconfigurable linear
optics~\cite{Ribeiro2016,Harris2018,Taballione2021}.

Recent advances in integrated photonics have enabled reconfigurable processors
containing tens to hundreds of tunable interferometric elements
~\cite{Harris2018,Perez2017,Wang2020,Morichetti2016,Bogaerts2020}. However, the
analog nature of these optical networks makes their performance highly
sensitive to fabrication imperfections, thermal drift, thermal crosstalk, and
environmental perturbations, which introduce phase and amplitude errors into
the implemented transformation
~\cite{Elmas2025,Milanizadeh2020,Bogaerts2020}. Phase instability in
programmable photonic processors has been modeled and experimentally
characterized, showing that uncontrolled drift can substantially degrade the
stability of interferometric signal processing
~\cite{Elmas2025,Litvin2025JLT}. Active multi-input phase stabilization using
on-chip feedback has also been demonstrated as a means of mitigating
time-dependent phase errors during processor operation
~\cite{Litvin2026Feedback}. As the mode count, circuit depth, and functional
complexity of programmable processors continue to increase, reliable, fast,
and automated calibration becomes a fundamental prerequisite for stable,
high-fidelity operation.

A widely used calibration approach is sequential node isolation, in which each
MZI is characterized individually after establishing a controlled optical path
through the surrounding mesh. Typically, one input of the target
interferometer is kept dark, while the preceding and downstream MZIs are
configured to provide a deterministic path from an external input to a
monitored output~\cite{Elmas2025,Perez2017}. Under these conditions, the
measured output power exhibits a sinusoidal dependence on the internal phase,
allowing the bar, cross, and intermediate operating points to be identified.
Alexiev \textit{et al.}~\cite{Alexiev2021}, for example, demonstrated
calibration of a rectangular interferometer mesh using external
photodetectors, while diagonal-routing architectures permit selected MZIs to be
accessed without accumulating phase errors from downstream
elements~\cite{Mojaver2023}.

The availability of such isolated paths depends strongly on the mesh
architecture. Under the independent-accessibility criterion considered in
Ref.~\cite{Mojaver2023}, only 14 of the 28 MZIs in an $8\times8$ Clements mesh
are independently accessible. Architectures such as the Bokun and Diamond
meshes provide independent access to every MZI, but require 40 and 49 MZIs,
respectively, compared with 28 in the corresponding Clements mesh, together
with additional auxiliary optical ports. These results expose a practical
trade-off between minimal optical depth and straightforward element-wise
access. Rather than modifying the processor architecture to create an isolated
diagonal path through every MZI, the method introduced here relaxes independent
accessibility as a calibration requirement by identifying the desired
operating points from an output-intensity variance signature even when both
local MZI inputs remain illuminated.

Self-configuring and parallel multi-input strategies have also been developed
to reduce the control burden associated with element-wise calibration.
Miller~\cite{Miller2013} introduced a progressive self-configuration framework
based on mutually orthogonal input modes, allowing interferometric elements to
be tuned using local feedback without reconstructing the complete transfer
matrix. Related self-configuring circuits have been experimentally
demonstrated for mode unscrambling, adaptive routing, and multifunctional
programmable photonics~\cite{Annoni2017,PerezLopez2020}. Similar concepts have
also been incorporated into adaptive and machine-learning-assisted photonic
processors~\cite{Shen2017,Hamerly2019}. These approaches are powerful, but may
require integrated monitoring, feedback control, mutually coherent multi-port
inputs, or carefully prepared training states. Maintaining these conditions in
large thermally tuned circuits remains challenging in the presence of
fabrication-induced phase mismatches, environmental drift, and time-dependent
phase noise.

An alternative multi-input calibration strategy based on temporal averaging
was introduced by Bandyopadhyay \textit{et al.}
~\cite{Bandyopadhyay2021}. In this method, the phase of an upstream
interferometer is swept over a complete $2\pi$ range and the transmission of
the target MZI is averaged over the sweep, thereby suppressing
phase-dependent contributions from an unwanted local input. The procedure
enables calibration in densely connected interferometric meshes and mitigates
errors caused by nonzero interfering fields. Nevertheless, each target element
requires an additional upstream phase sweep, increasing the number of
measurements and introducing sensitivity to drift and actuator nonlinearity
when the applied sweep does not accurately cover a complete phase period
~\cite{Bandyopadhyay2021}.

Other calibration approaches employ additional hardware or more extensive
signal processing. Xu \textit{et al.}~\cite{Xu2022} demonstrated a
self-calibrating programmable photonic circuit based on on-chip reference
paths and Kramers--Kronig phase retrieval, while Milanizadeh
\textit{et al.}~\cite{Milanizadeh2020} developed control methods based on
thermal-eigenmode decomposition and signal labeling for suppressing thermal
crosstalk. Auto-calibrating circuits designed for defect resilience and global
strategies that learn nonlinear phase--current relations across many tuning
elements have also been proposed~\cite{Markowitz2023,Zheng2024}. Related
approaches based on direct-feedback optimization, gradient-based tuning, and
stochastic control have been explored for large interferometric systems
~\cite{Hamerly2019,Perez2017,Wang2020}. The influence of calibration choices
on robustness and energy consumption has also been investigated in
reconfigurable photonic processors~\cite{Litvin2025}. Although these methods provide
powerful mechanisms for compensating device imperfections, they may increase
the required hardware, control complexity, calibration time, or computational
overhead.

Despite these advances, existing calibration strategies generally rely on
some combination of strict node isolation, independently accessible routing
paths, coherent multi-port excitation, calibrated auxiliary phase sweeps,
reference channels, embedded monitoring, global numerical optimization, or
detailed phase--voltage characterization
~\cite{Mojaver2023,Bandyopadhyay2021,Xu2022,Milanizadeh2020,
PerezLopez2020,Markowitz2023,Zheng2024}. These requirements become
progressively more demanding as the number of tunable elements and the circuit
depth increase. There is therefore a need for experimentally simple,
intensity-only calibration observables whose extrema remain identifiable
without an absolute phase reference, conventional node isolation, or a
calibrated actuator response.

In this work, we introduce and experimentally demonstrate a local
variance-based calibration method for programmable photonic interferometer
meshes. The method identifies operating points from the statistical
fluctuations of a selected output intensity under controlled phase
perturbations. For MZI calibration, the output standard deviation follows a
characteristic $|\sin\theta|$ dependence, whose minima identify the bar and
cross states. For standalone phase-shifter calibration under balanced
interference, the corresponding observable follows a $|\cos\phi|$ dependence,
with minima at the quadrature operating points. Since the standard deviation is
the square root of the variance, both observables possess identical
minimizers. The method therefore requires only intensity measurements and a
reproducible ensemble of perturbation settings.

A central practical feature of the approach is that the perturbation settings
need not represent calibrated phases or a uniform $0$--$2\pi$ sweep. For MZI
calibration, an arbitrary nondegenerate finite ensemble changes only the
multiplicative scale of the variance curve and leaves the locations of the bar
and cross minima unchanged. This enables the same small set of random voltage
values to be reused throughout the mesh without prior phase--voltage
characterization. The method consequently provides a complementary solution
to architectures designed for complete independent accessibility: it retains
the compact Clements topology while replacing the isolated transmission
minimum by a statistical calibration signature.

We validate the method experimentally on an $8\times8$ silicon-nitride
programmable photonic processor using an automated two-stage procedure.
Starting from arbitrary initial settings, all MZIs are calibrated first,
followed by the standalone phase shifters under balanced-interference
conditions. The procedure is applied to the unmodified 28-MZI Clements mesh,
including elements that do not possess independently accessible diagonal
paths under the conventional dark-input criterion. We subsequently configure
an embedded $4\times4$ Hadamard transformation using a Clements decomposition
~\cite{Clements2016} and compare the measured output response with that
obtained using the manufacturer-provided calibration. Under the experimental
conditions considered here, the variance-based calibration produces more
uniform and temporally stable output signals, demonstrating the practical
potential of local output statistics as a low-overhead calibration observable
for programmable photonic processors.

\section{Variance-Based Calibration Principle}

We introduce a local variance-based calibration principle for programmable
photonic interferometer meshes. The central idea is to replace phase-resolved
fringe fitting by a statistical observable: the variance of an output intensity
measured under controlled phase perturbations. Because the locations of the
variance minima are fixed by the interferometric transfer function, calibration
can be performed using intensity-only measurements and without prior knowledge
of the phase--voltage response of the thermo-optic actuators.

The method is local in the sense that each tunable element is calibrated from
the statistical response of a selected output channel. It does not require
conventional node isolation, reference paths, phase-resolved detection, or
globally orthogonal training fields. The same principle yields two complementary
calibration signatures: a $|\sin\theta|$ dependence for Mach--Zehnder
interferometers (MZIs), and a $|\cos\phi|$ dependence for phase shifters (PSs)
under balanced interference. These two cases are derived below.

\begin{figure}[H]
    \centering
        \includegraphics[width=0.7\textwidth]{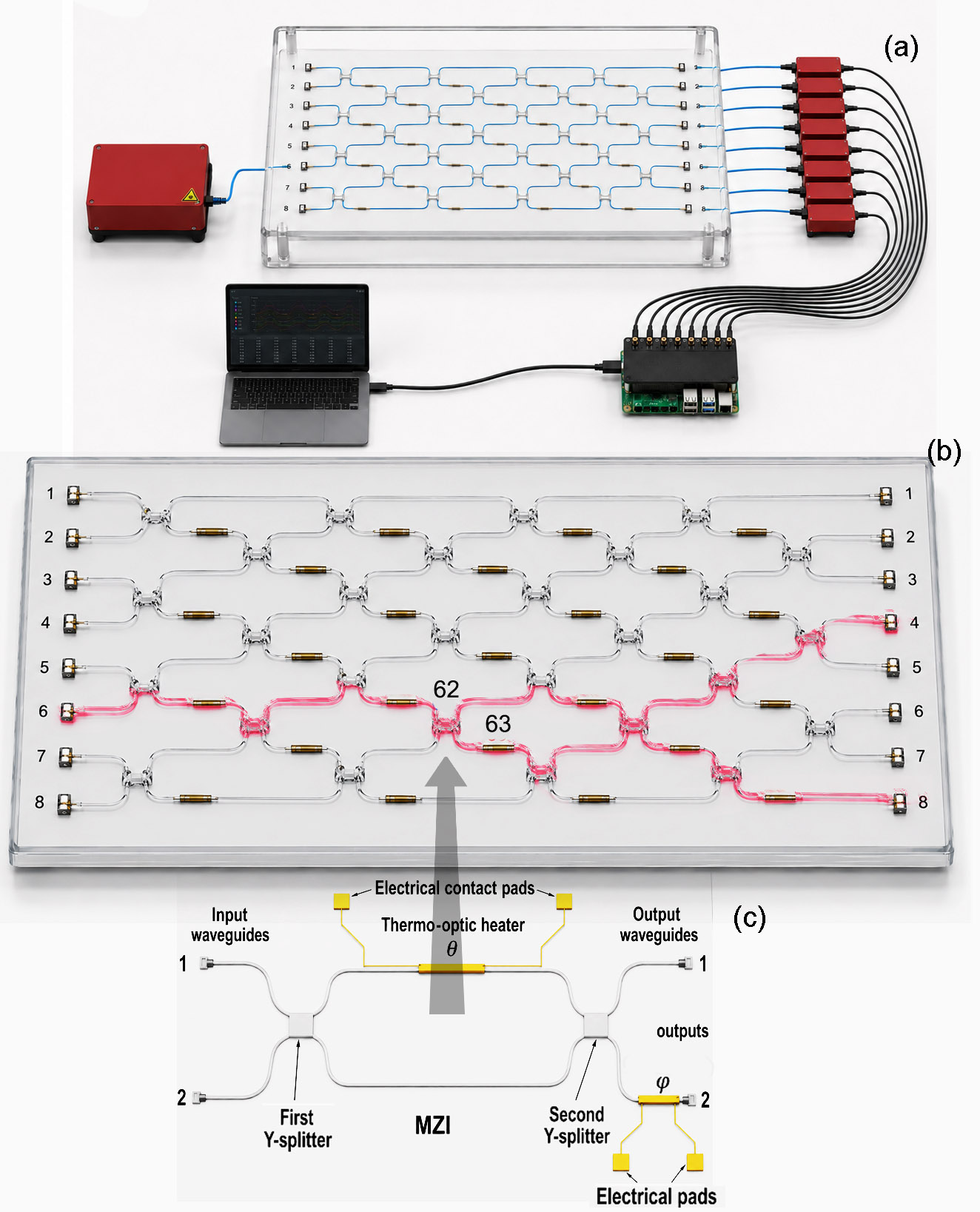}
        \caption{
    Experimental platform and programmable photonic processor.
    \textbf{(a)} Experimental arrangement used for intensity-based
    characterization of the eight-mode programmable photonic processor.
    Light from a continuous-wave laser is coupled into a selected input port
    through a polarization-maintaining fiber. The eight output ports are
    connected to individual photodetectors, whose electrical signals are
    digitized by an eight-channel analog-to-digital conversion stage and
    recorded on a computer. Blue lines denote optical fiber connections,
    whereas black lines denote electrical connections.
    \textbf{(b)} Rendered layout of the \(8\times 8\) interferometric mesh,
    showing the arrangement of tunable Mach--Zehnder interferometers and
    standalone phase shifters. The representative MZI and phase shifter used
    in the calibration procedure are marked by 62 and 63, respectively.
    \textbf{(c)} Enlarged visualization of a single tunable unit cell. The
    unit cell consists of an MZI with internal thermo-optic phase shift
    \(\theta\), followed by an output phase shifter with phase \(\phi\).
    Gold structures represent the thermo-optic heaters and their electrical
    contacts.
    }

    \label{fig:mesh_layout}
\end{figure}

\subsection{Unit-Cell Transfer Matrix}

Each tunable unit cell consists of a Mach--Zehnder interferometer with
internal phase shift $\theta$, followed by a phase shifter with phase
$\phi$ on one output arm. Under the convention adopted in this work, the
corresponding transfer matrix is

\begin{equation}
\mathbf{U}(\theta,\phi)
=
\frac{1}{2}
\begin{pmatrix}
1-e^{-i\theta}
&
-i\left(1+e^{-i\theta}\right)
\\[4pt]
-i\left(1+e^{-i\theta}\right)e^{-i\phi}
&
-\left(1-e^{-i\theta}\right)e^{-i\phi}
\end{pmatrix}.
\label{eq:unit_cell_matrix}
\end{equation}

The internal phase $\theta$ controls the effective splitting ratio of the
MZI, whereas $\phi$ introduces an additional relative phase on the second
output arm. For $\theta=0$ and $\theta=\pi$, the MZI operates at the bar
and cross states, respectively, while $\theta=\pi/2$ corresponds to
balanced splitting.

Detailed derivations of the MZI and phase-shifter calibration
relations are provided in Supplementary Sections~\ref{supp:mzi_derivation}
and~\ref{supp:ps_derivation}, respectively.
\subsection{Variance-Based Mach--Zehnder Interferometer Calibration}
\label{sec:mzi_calibration}

Each unit cell contains a tunable Mach--Zehnder interferometer followed by an
output phase shifter. The internal MZI phase is denoted by $\theta$. In the
ideal case, $\theta=0$, $\pi/2$, and $\pi$ correspond to the bar, balanced
50/50, and cross states, respectively. At the bar and cross states, the
monitored output intensity becomes insensitive to variations of the relative
phase between the two input fields.

To exploit this property, coherent fields are injected into the two input ports
of the target MZI with amplitudes $A e^{i\phi}$ and $B$, where $A,B>0$ and
$\phi$ denotes the controllable relative input phase. For each fixed trial
value of $\theta$, the output intensity is recorded for several values of
$\phi$, and its empirical standard deviation is evaluated as a function of
$\theta$. For the unit-cell convention adopted here, the monitored output
intensity is
\begin{equation}
P_1(\theta,\phi)
=
\frac{A^2+B^2}{2}
+
\frac{B^2-A^2}{2}\cos\theta
-
AB\sin\theta\cos\phi .
\label{eq:mzi_power_main}
\end{equation}

Averaging over a uniformly distributed relative phase
$\phi\in[0,2\pi)$ gives
\begin{equation}
\sigma(\theta)
=
\frac{AB}{\sqrt{2}}|\sin\theta|.
\label{eq:MZI_sigma_main}
\end{equation}

\begin{figure}[H]
    \centering
    \begin{overpic}[width=0.47\textwidth]{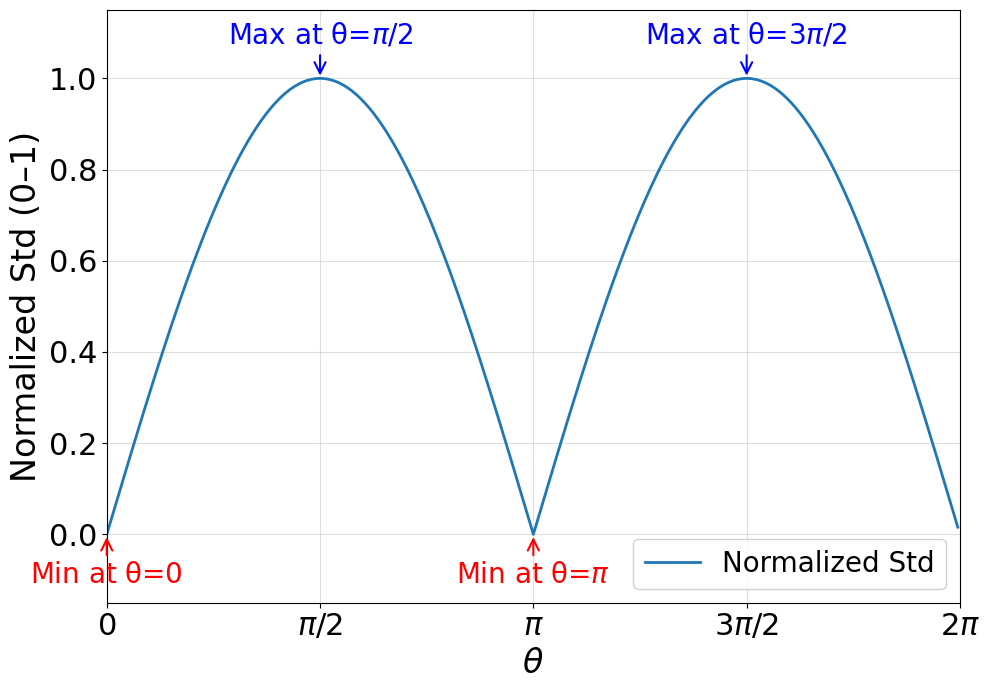}
        \put(100,70){\small(a)}
    \end{overpic}
    \hspace{0.6em}
    \begin{overpic}[width=0.47\textwidth]{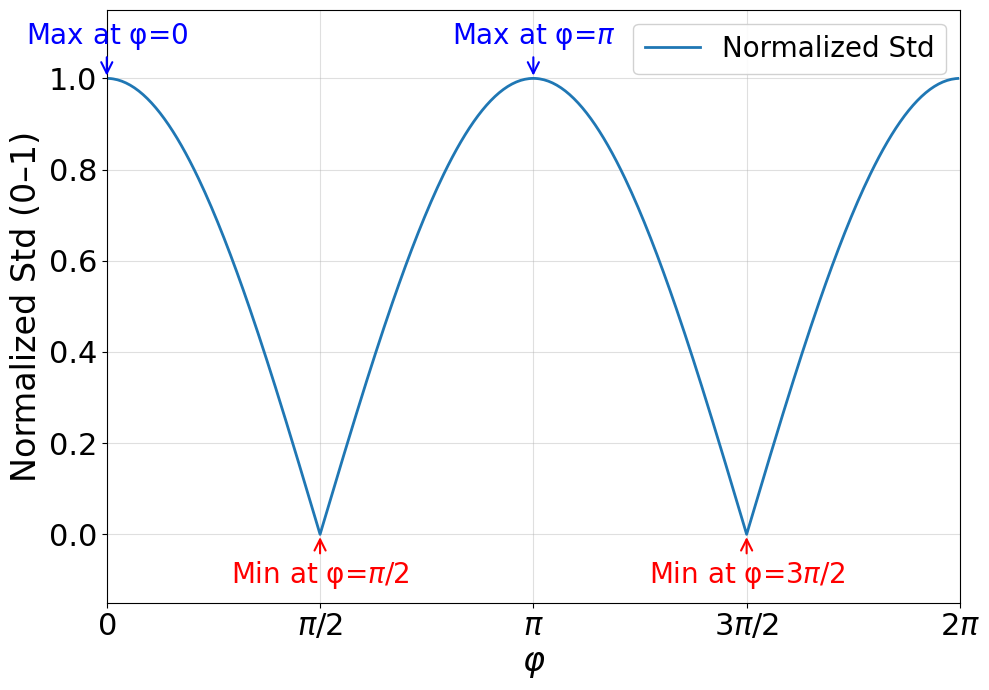}
        \put(100,70){\small(b)}
    \end{overpic}

    \vspace{0.3cm}

    \caption{
    Theoretical variance-based calibration signatures.
    \textbf{(a)} Normalized standard deviation of the monitored MZI output
    intensity as a function of the internal phase $\theta$, showing the
    characteristic $|\sin\theta|$ dependence. The minima at
    $\theta=0$ and $\theta=\pi$ identify the bar and cross operating points,
    respectively.
    \textbf{(b)} Normalized standard deviation of the phase-shifter
    calibration signal as a function of the phase-shifter phase $\phi$ under
    balanced-interference conditions, showing the characteristic
    $|\cos\phi|$ dependence. The minima at $\phi=\pi/2$ and
    $\phi=3\pi/2$ identify the quadrature operating points.
    }
    \label{fig:theoretical_signatures}
\end{figure}

The resulting theoretical calibration signature is shown in
Fig.~\ref{fig:theoretical_signatures}(a). The output standard deviation
vanishes at $\theta=0$ and $\theta=\pi$, corresponding to the bar and cross
states, respectively, and reaches its maximum at the balanced-splitting
points $\theta=\pi/2$ and $\theta=3\pi/2$. MZI calibration therefore reduces
to identifying the minima of the measured output standard deviation.

In the experiments, we use the empirical standard deviation
$\hat{\sigma}$ as the calibration observable. Since $\hat{\sigma}$ is the
square root of the empirical variance, the variance and standard deviation
have identical minimizers. We therefore retain the term variance-based
calibration while reporting and plotting the corresponding standard
deviation.

\paragraph{Finite random phase sampling.}
A key practical feature of the method is that a calibrated continuous
$0$--$2\pi$ phase sweep is not required. Instead, one may use a fixed finite
ensemble of random phase settings,
\begin{equation}
\Phi_M=\{\phi_1,\phi_2,\ldots,\phi_M\}, \qquad \phi_k\in[0,2\pi).
\end{equation}
The empirical standard deviation evaluated over this finite ensemble is
\begin{equation}
\sigma^{(M)}(\theta)
=
AB|\sin\theta|\sqrt{C_M},
\label{eq:mzi_finite_sigma}
\end{equation}
where
\begin{equation}
C_M
=
\frac{1}{M}\sum_{k=1}^{M}\cos^2\phi_k
-
\left(
\frac{1}{M}\sum_{k=1}^{M}\cos\phi_k
\right)^2 .
\label{eq:CM_main}
\end{equation}
The factor $C_M$ depends only on the chosen phase ensemble and is independent
of $\theta$. We emphasize that neither the input amplitudes $A$ and $B$ nor the
absolute values of the applied phases $\phi_k$ need to be known. It is sufficient
that both input amplitudes remain nonzero and approximately constant during the
calibration scan, and that the same nondegenerate set of phase settings
$\{\phi_k\}_{k=1}^{M}$ can be reproducibly applied at every trial value of
$\theta$. The unknown amplitudes and phase ensemble determine only the
multiplicative factor $AB\sqrt{C_M}$ and therefore do not shift the locations of
the variance minima. Consequently, finite random sampling rescales the variance
curve while preserving the bar and cross operating points, even when the applied
voltage values are not calibrated phases and do not form a uniform sweep.

For independent uniformly distributed phase samples,
$\mathbb{E}[C_M]=(M-1)/(2M)$, which approaches $1/2$ as $M$ increases. More
importantly, $C_M$ is nonzero with probability one for a nondegenerate random
ensemble, so the characteristic $|\sin\theta|$ dependence is preserved. This is
what allows the same small set of random voltage values to be reused throughout
the mesh without prior phase--voltage characterization.

In practice, we find that $M=5$ random phase samples per MZI are sufficient
to locate the variance minimum reliably. This substantially reduces the control
overhead compared with full phase sweeps and avoids the need for prior
phase--voltage characterization.
\subsection{Variance-Based Phase-Shifter Calibration}
\label{sec:ps_calibration}

The same variance principle can be used to calibrate a phase shifter. In this
case, the phase shifter under test controls the relative phase $\phi$ between
two fields that interfere in a balanced MZI. The internal MZI phase $\theta$ is
then sampled or scanned, and the output-intensity variance with respect to
$\theta$ is used as the calibration observable.

Using the same output-power expression in
Eq.~\eqref{eq:mzi_power_main}, the variance with respect to $\theta$ at fixed
$\phi$ is
\begin{equation}
\mathrm{Var}_{\theta}[P_1(\cdot,\phi)]
=
\frac{(A^2-B^2)^2}{8}
+
\frac{A^2B^2}{2}\cos^2\phi .
\label{eq:var_PSI}
\end{equation}

For balanced inputs, $A=B$, this reduces to
\begin{equation}
\sigma(\phi)
=
\frac{A^2}{\sqrt{2}}|\cos\phi|.
\label{eq:PS_sigma_main}
\end{equation}

The resulting theoretical calibration signature is shown in
Fig.~\ref{fig:theoretical_signatures}(b). The standard deviation reaches its
maximum at $\phi=0$ and $\phi=\pi$, and vanishes at the quadrature points
$\phi=\pi/2$ and $\phi=3\pi/2$. These minima identify the calibrated
phase-shifter operating points. At quadrature, the monitored output intensity
is first-order insensitive to fluctuations or mis-setting of the sampled
internal MZI phase $\theta$.

\paragraph{Finite random sampling of the auxiliary phase.}
As for the MZI calibration, the auxiliary phase $\theta$ need not be swept
continuously. A fixed finite set
\begin{equation}
\Theta_M=\{\theta_1,\theta_2,\ldots,\theta_M\}, \qquad \theta_k\in[0,2\pi),
\end{equation}
can be used to estimate the variance. The resulting empirical standard
deviation retains the same $|\cos\phi|$ dependence, again up to a scale factor
that depends only on the chosen sample set. Therefore, the quadrature points of
the phase shifter can be identified from variance minima without requiring an
absolute phase reference or a calibrated phase--voltage map.

Thus, PS calibration is achieved by tuning the phase shifter under test until
the measured variance with respect to the auxiliary MZI phase is minimized. The
two possible minima, $\pi/2$ and $3\pi/2$, are equivalent for quadrature
calibration; distinguishing the branch is only necessary if an absolute phase
assignment is required.

\subsection{Calibration Order and Scaling}

Calibration begins from arbitrary and unknown phase settings and proceeds
layer by layer from the output side of the mesh toward the inputs. After
the interferometers in a given layer have been calibrated, they are fixed
at known bar or cross states and subsequently provide deterministic
readout paths for the calibration of the preceding layer.

The variance observable is separable across the non-overlapping
interferometers of a single Clements layer. Provided that the local
input fields at each interferometer are nonzero, the applied phase
ensembles are nondegenerate and reproducible, and the corresponding
outputs are independently resolved, all interferometers within the
layer can be calibrated concurrently. Unlike parallel-nullification
methods that require specifically calculated multiport input vectors,
the variance minima are independent of the unknown local input
amplitudes and of the absolute applied phase values.

An $N$-mode rectangular Clements mesh contains $N(N-1)/2$
interferometers arranged in $\mathcal{O}(N)$ layers. For a fixed
number of trial voltages and phase realizations, layer-wise
parallelization can therefore reduce the number of sequential
measurement rounds for the MZI-calibration stage from
$\mathcal{O}(N^2)$ to $\mathcal{O}(N)$, corresponding to an ideal
speed-up of approximately $N/2$. 

\section{Experimental Platform and Automated Calibration Procedure}

\subsection{Photonic Processor Layout and Measurement Setup}

Figure~\ref{fig:mesh_layout} shows the layout of the programmable 8$\times$8 photonic interferometer mesh used in this work. The device is an 8-mode variant of the universal silicon-nitride photonic processor architecture previously reported in~\cite{Taballione2021}, which was originally demonstrated in a 12-mode configuration. The processor is based on stoichiometric silicon nitride (Si$_3$N$_4$) waveguides fabricated in the TriPleX platform and operates at a wavelength of 1550~nm.

The interferometer consists of a square mesh of tunable Mach--Zehnder interferometers (MZIs) arranged in 8 layers, enabling full all-to-all mode coupling. Each unit cell comprises a tunable beam splitter implemented as an MZI with an internal thermo-optic phase shifter, followed by an external output phase shifter. The thermo-optic phase shifters are realized using resistive platinum heaters with typical $\pi$-phase voltages on the order of 10~V.
Light is coupled into and out of the chip using polarization-maintaining fiber arrays, and the device is mounted inside an actively temperature-stabilized control enclosure. All phase shifters are addressed through multi-channel digital-to-analog converters and controlled via a fully automated software interface. Output optical powers are measured using InGaAs photodiodes connected to a multi-channel data acquisition system.

\subsection{Experimental Calibration of Mach--Zehnder Interferometers}

The variance-based MZI calibration procedure described in
Section~\ref{sec:mzi_calibration}  is applied sequentially to all interferometers in the $8\times8$ mesh, starting from the output layer and proceeding layer by layer toward the input p
orts. At the beginning of this process, all MZIs and phase shifters are in arbitrary and generally unknown phase states. As calibration proceeds, each newly calibrated interferometer is fixed to its identified bar or cross operating point and subsequently serves as a stabilized routing element for the calibration of preceding layers.

For each target MZI, the internal phase $\theta$ is scanned over its available tuning range by varying the corresponding heater voltage. Since no prior phase--voltage calibration is assumed for the input phase shifters at this stage, a full $0$--$2\pi$ phase sweep is not performed. Instead, a fixed set of $M=5$ random phase-shifter voltage values is generated once and reused throughout the entire calibration procedure. These same five random phase settings are applied at each trial value of $\theta$ to generate a discrete ensemble $\{\phi_k\}$ of relative input phases. While in theory as few as two phase realizations are sufficient to produce a nonzero variance signal, using $M=5$ samples provides improved statistical robustness and smoother variance curves in the experimental data.

For each of the $M$ phase realizations, the optical output power at a single output port of the target MZI is recorded using an InGaAs photodiode, yielding a set of $M$ intensity samples for each $\theta$. From these measurements, the empirical output variance is computed and normalized, and the value of $\theta$ minimizing the variance is identified as the calibrated bar or cross operating point.

\begin{figure}[t]
    \centering
    \begin{overpic}[width=0.43\textwidth]{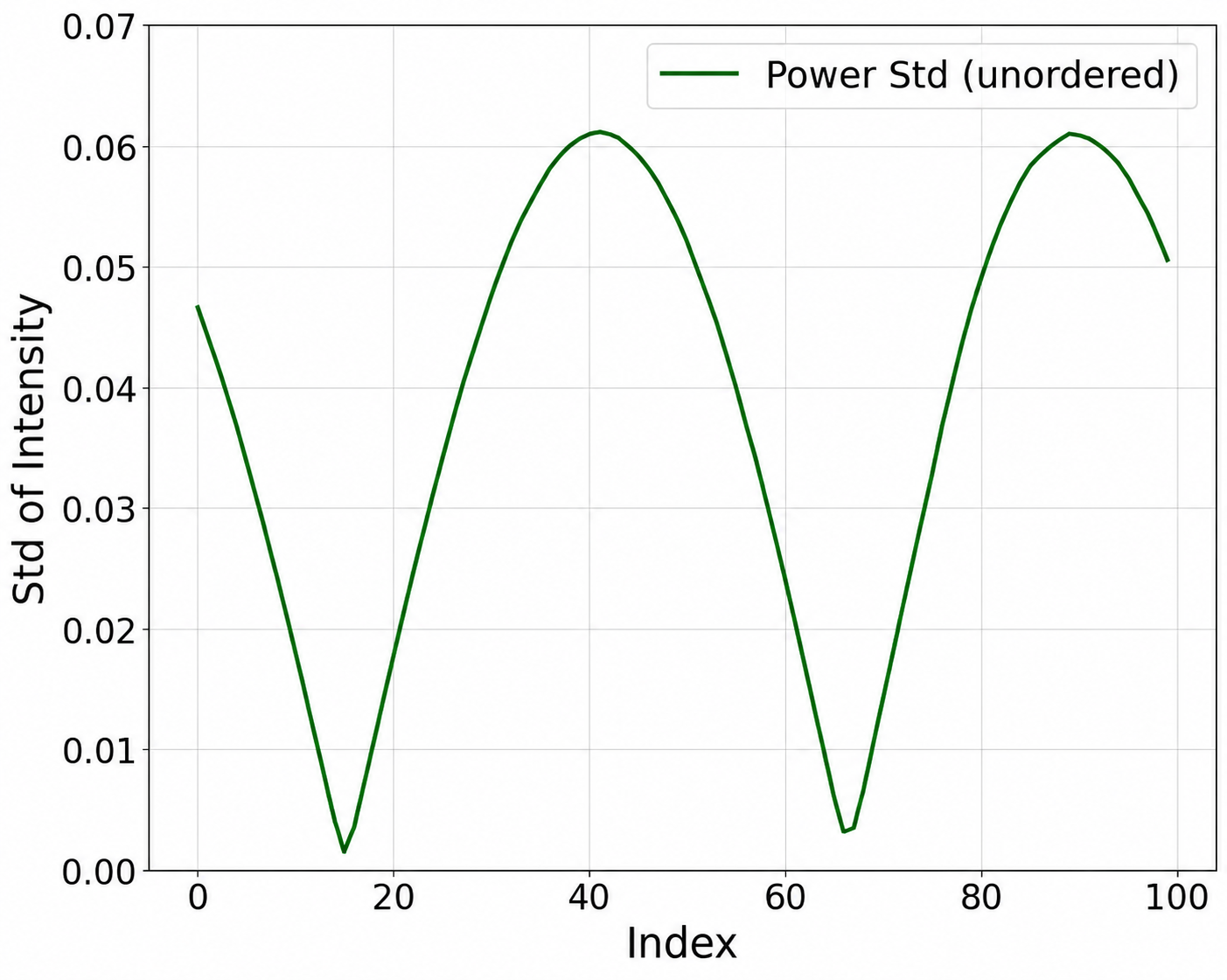}
        \put(100,70){\small(a)}
    \end{overpic}
    \hspace{0.6em}
    \begin{overpic}[width=0.43\textwidth]{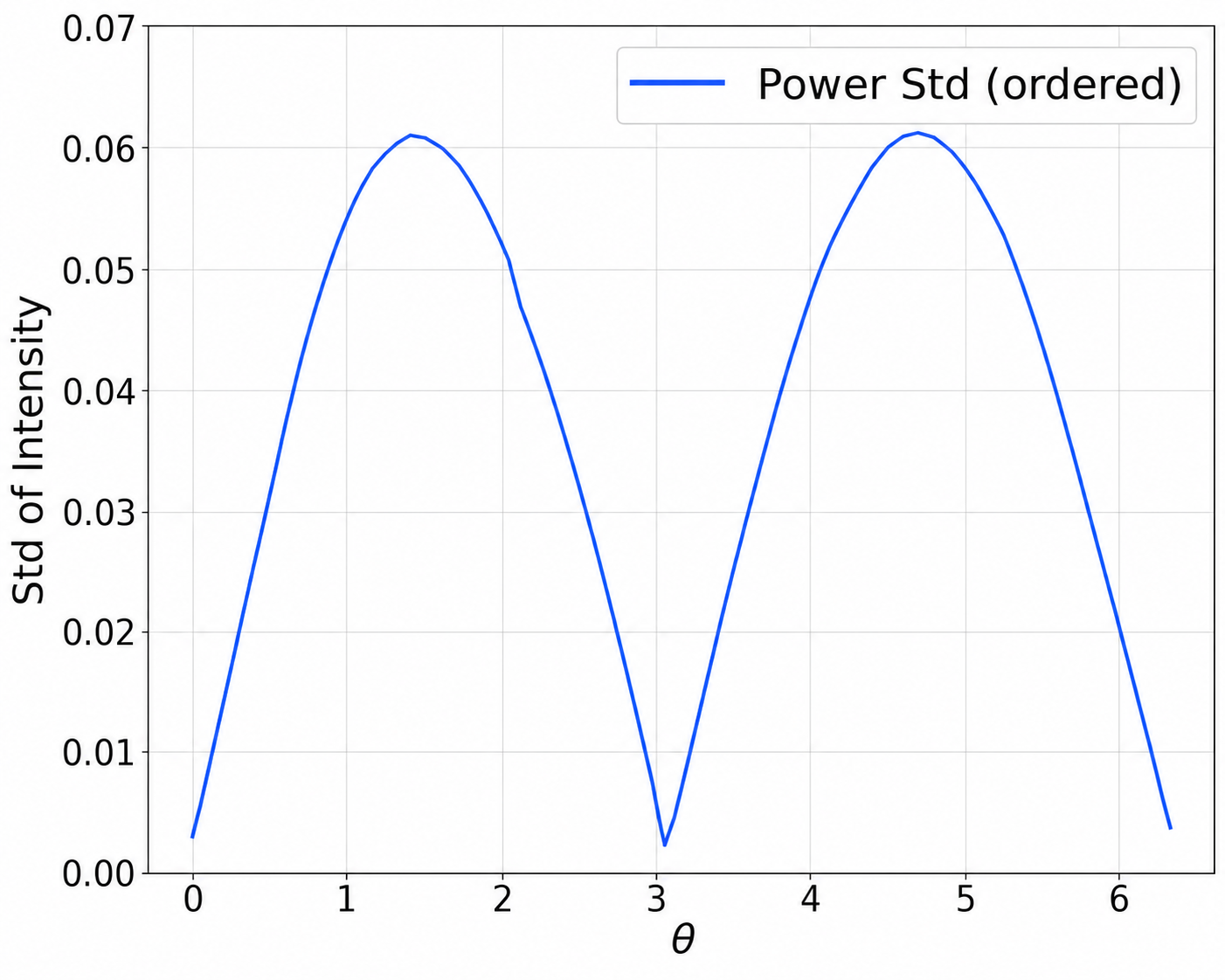}
        \put(102,70){\small(b)}
    \end{overpic}

    \vspace{0.3cm}

    \begin{overpic}[width=0.43\textwidth]{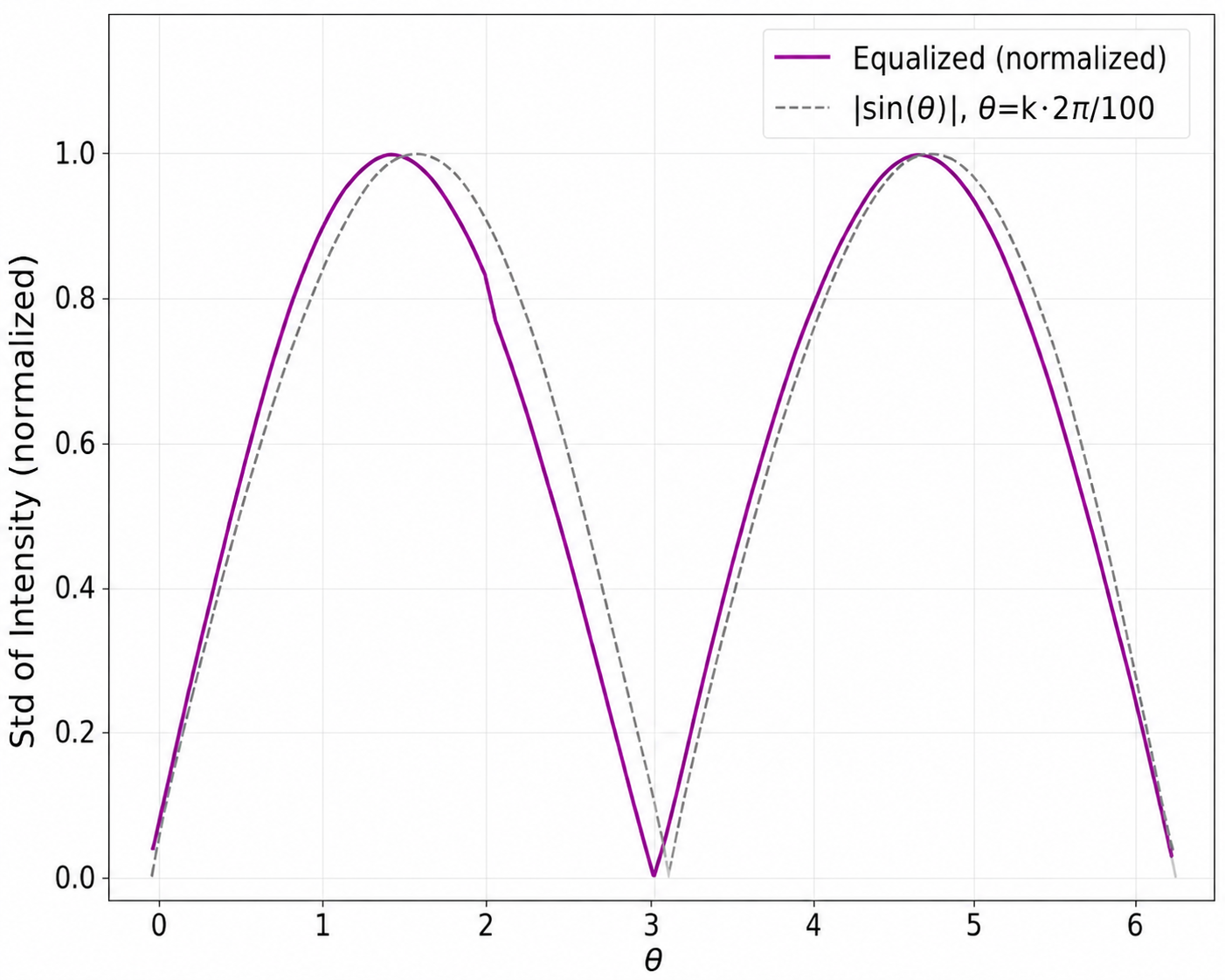}
        \put(100,70){\small(c)}
    \end{overpic}
    \hspace{0.6em}
    \begin{overpic}[width=0.43\textwidth]{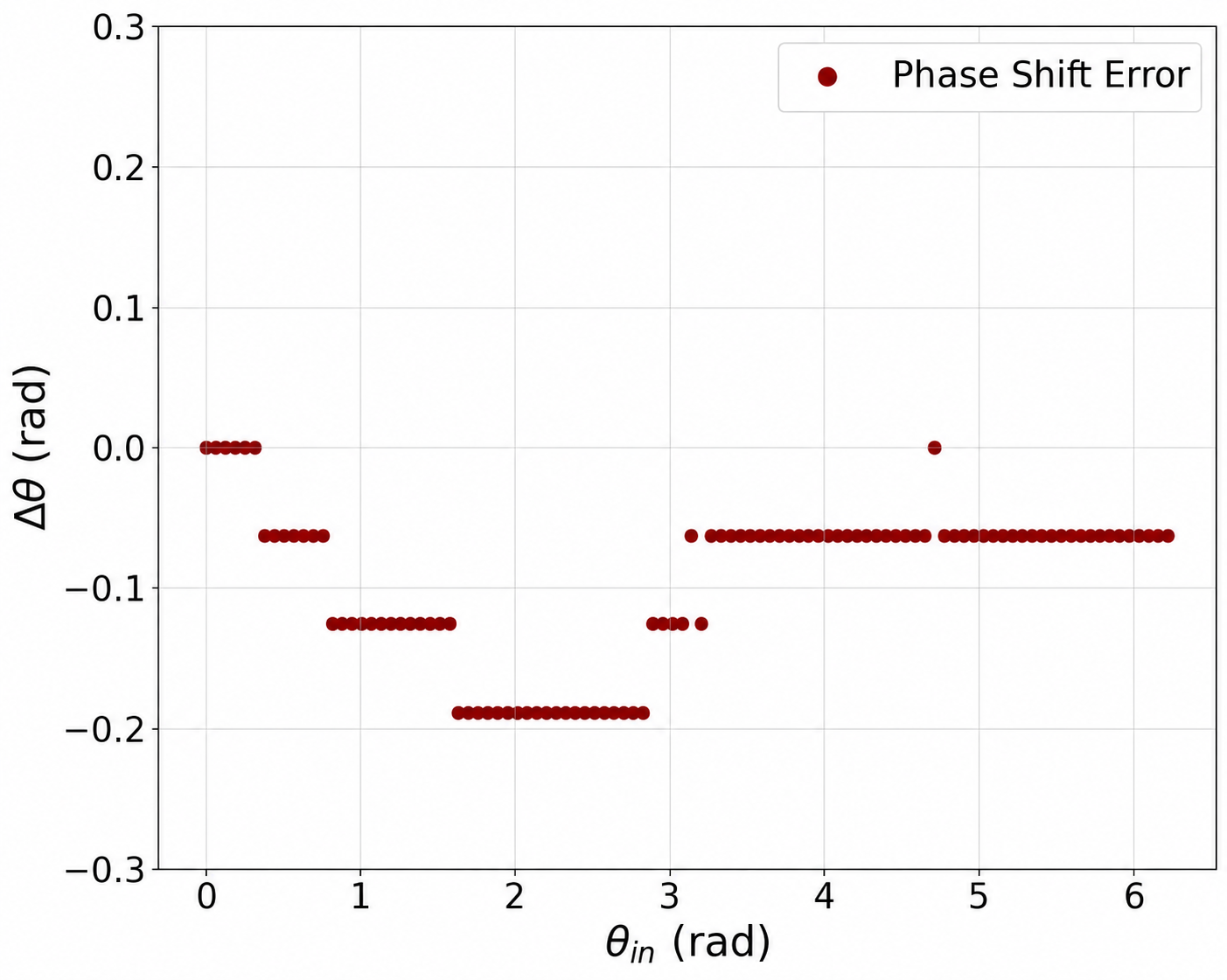}
        \put(102,70){\small(d)}
    \end{overpic}

    \caption{Experimental variance-based MZI calibration for MZI$_{62}$.
    (a) Raw unnormalized output variance as a function of heater phase $\theta$ (unordered).
    (b) Same data after ordering.
    (c) Ordered and normalized variance following the theoretical $|\sin\theta|$ dependence.
    (d) Extracted residual phase error $\Delta\theta$ after calibration.}
    \label{fig:MZI_calibration}
\end{figure}

Figure~\ref{fig:MZI_calibration} shows a representative experimental calibration result for MZI$_{62}$ located in a central layer of the mesh. Figures~\ref{fig:MZI_calibration}(a) and (b) show the raw unnormalized and ordered output variance as a function of the heater phase $\theta$, respectively. The corresponding normalized variance is shown in Fig.~\ref{fig:MZI_calibration}(c). The measured curve exhibits the characteristic $|\sin\theta|$ dependence predicted by Eq.~\eqref{eq:MZI_sigma_main}, despite discretely sampled phase values, uncontrolled upstream interference, and experimental noise. A clear and well-defined variance minimum is observed, identifying the calibrated bar/cross operating point of the interferometer.

After fixing the heater voltage corresponding to this variance minimum, the residual phase error of the calibrated MZI is evaluated. Figure~\ref{fig:MZI_calibration}(d) shows the extracted phase deviation $\Delta\theta$ as a function of the applied heater phase $\theta_{\mathrm{in}}$ for this representative device. The residual phase error remains within approximately $\pm0.2$~rad across the full tuning range, confirming reliable convergence of the variance-based MZI calibration procedure. All MZIs across the processor exhibit comparable behavior.

\subsection{Experimental Calibration of Phase Shifters}

After completion of the variance-based calibration of all Mach--Zehnder interferometers, the standalone output phase shifters are calibrated using the protocol described in Section~\ref{sec:ps_calibration}. As predicted by Eq~\eqref{eq:var_PSI},
the characteristic trigonometric dependence of the variance on the relative input phase $\phi$ is obtained only under balanced input amplitudes $A=B$.

To experimentally satisfy this condition, the MZI directly preceding each phase shifter is set to its 50/50 operating point ($\theta=\pi/2$), thereby ensuring equal optical power in the two interfering arms. This balanced-interference configuration is maintained throughout the phase-shifter calibration stage. The phase shifters are then calibrated sequentially following a diagonal order across the mesh. This ordering is chosen because the uppermost waveguide does not contain a tunable phase shifter and is fixed to zero phase by design. The diagonal sequence guarantees that the required balanced condition is preserved and that all previously calibrated elements remain stabilized.

For each target phase shifter, the internal phase parameter $\theta$ of the corresponding unit cell is sampled using a fixed set of random heater voltage values, while the relative input phase $\phi$ is swept. As in the MZI calibration stage, no prior phase--voltage characterization is assumed. Instead, a discrete ensemble of random voltage values is reused throughout the calibration procedure to probe the statistical response. For each value of $\phi$, the output optical power at a single output port is recorded and the empirical variance with respect to $\theta$ is computed.

\begin{figure}[t]
    \centering
    \begin{overpic}[width=0.40\textwidth]{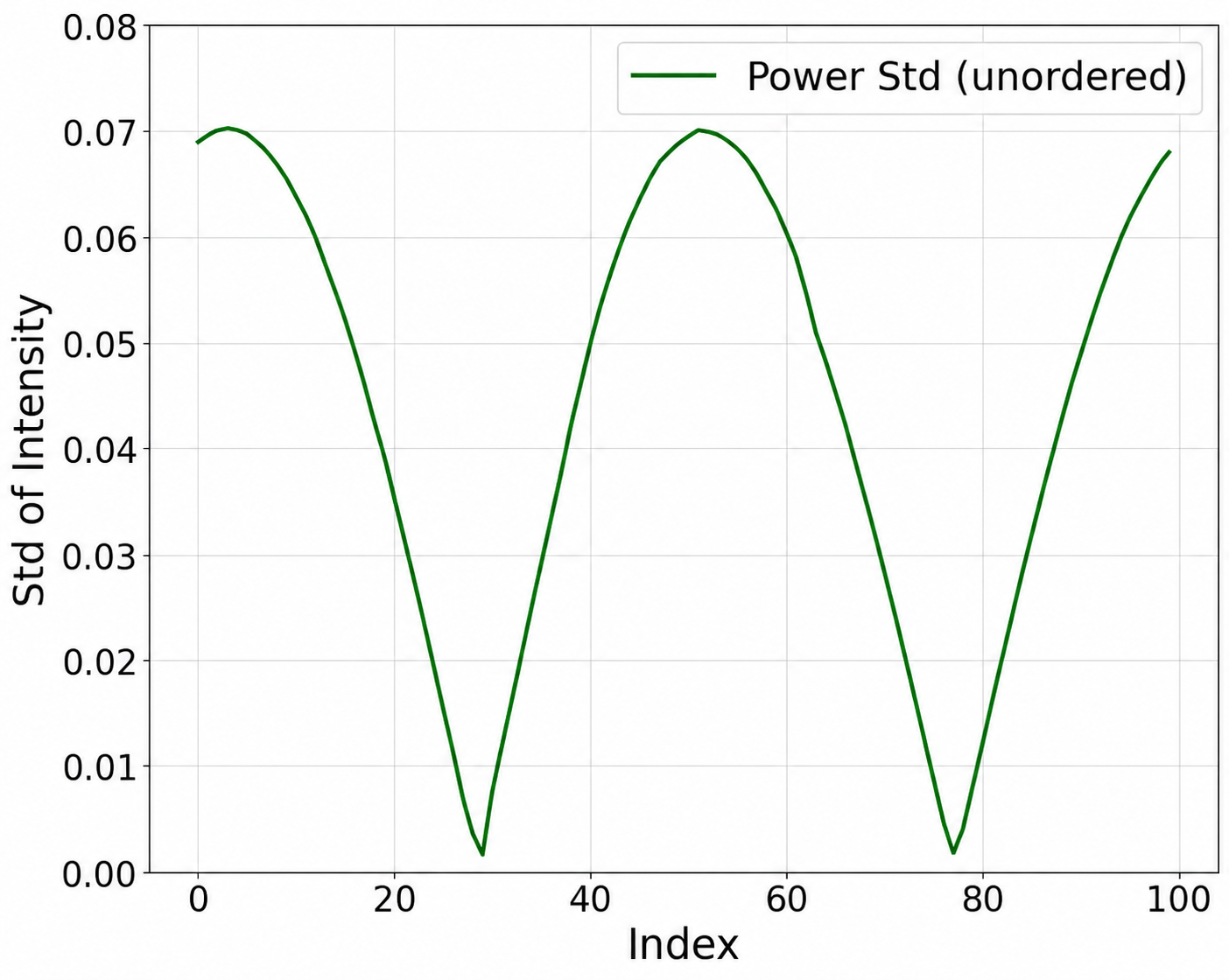}
        \put(102,70){\small(a)}
    \end{overpic}
    \hspace{0.6em}
    \begin{overpic}[width=0.40\textwidth]{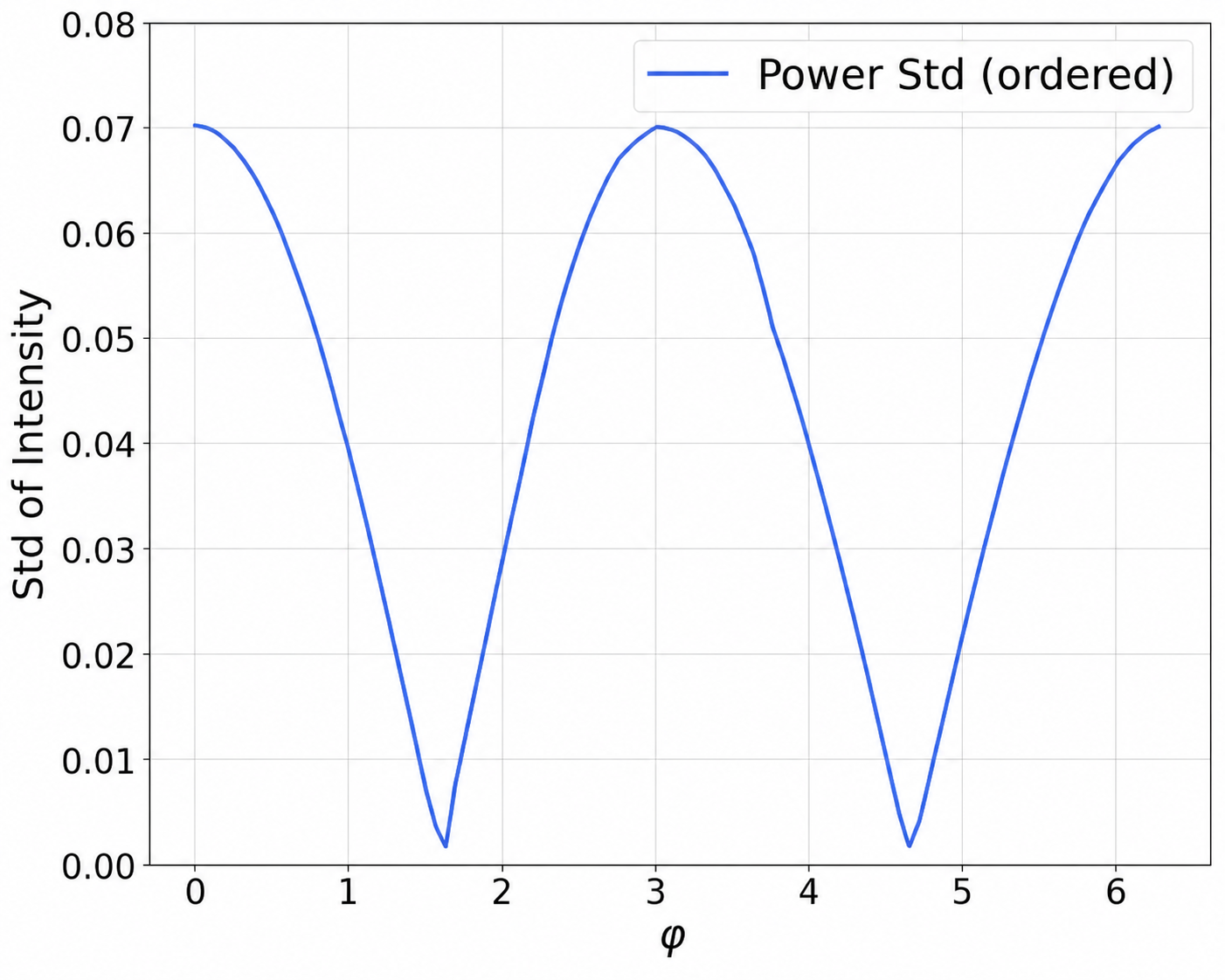}
        \put(102,70){\small(b)}
    \end{overpic}

    \vspace{0.3cm}

    \begin{overpic}[width=0.40\textwidth]{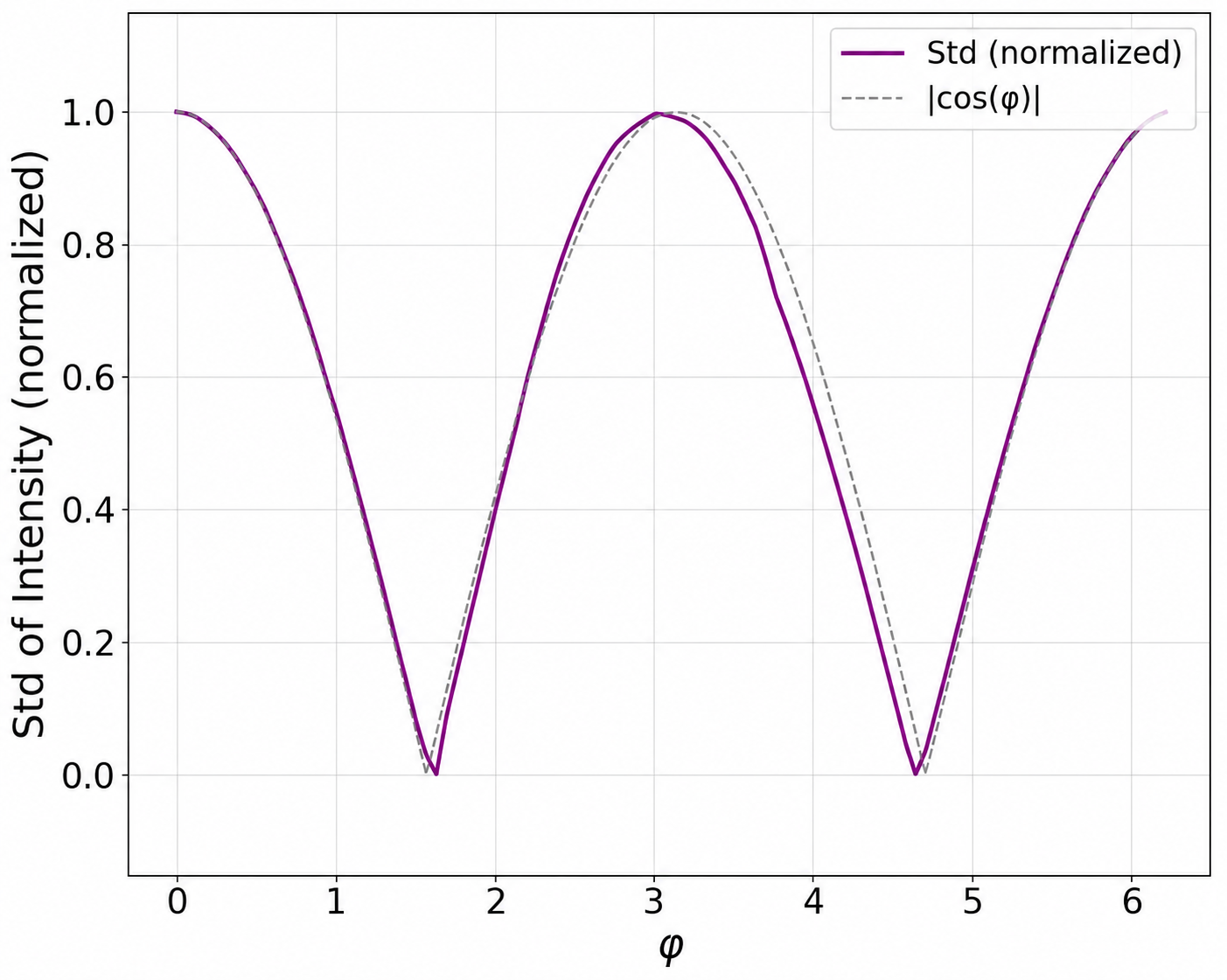}
        \put(102,70){\small(c)}
    \end{overpic}
    \hspace{0.6em}
    \begin{overpic}[width=0.40\textwidth]{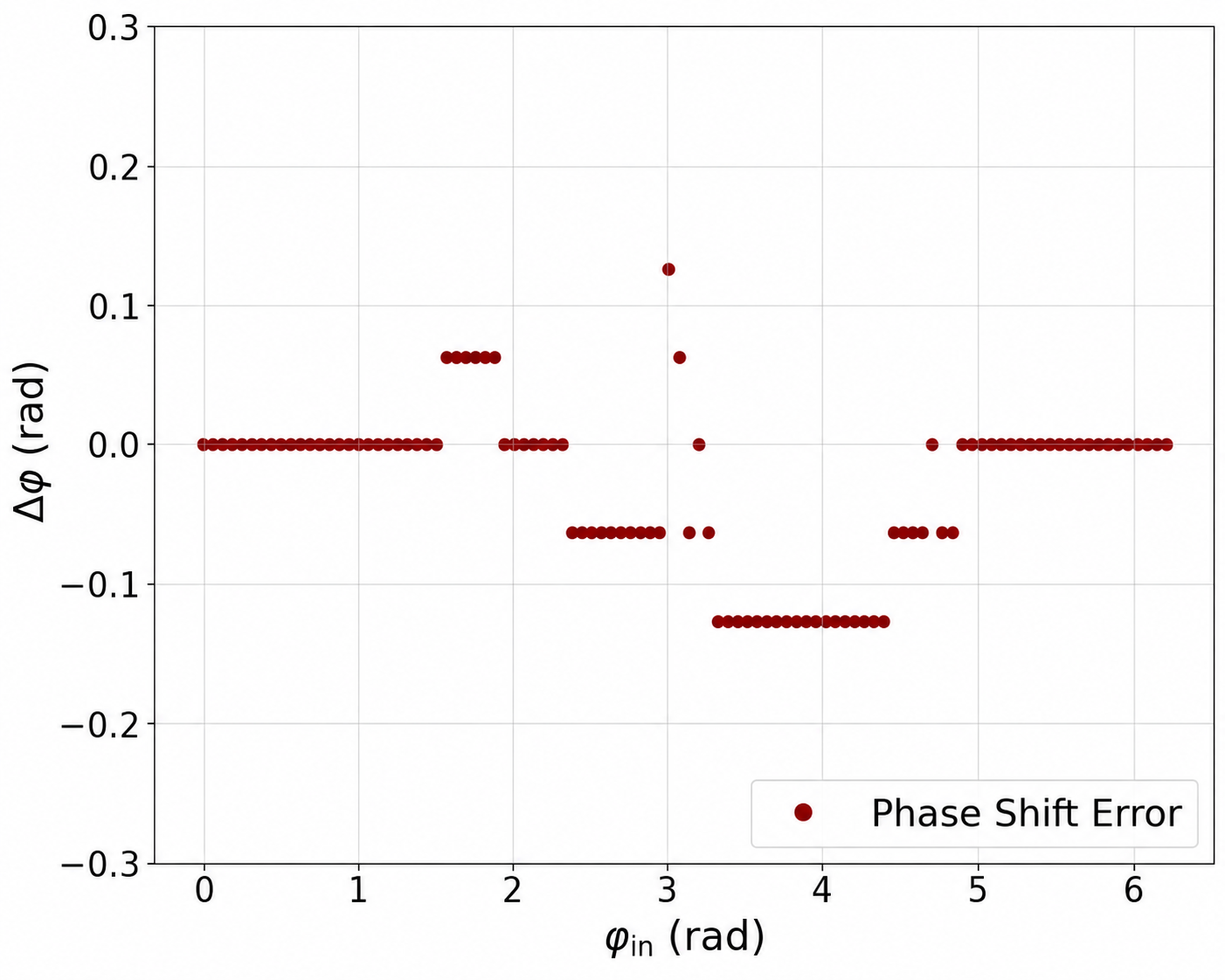}
        \put(102,70){\small(d)}
    \end{overpic}

    \caption{Experimental variance-based PS calibration for PS$_{63}$.
    (a) Raw unnormalized output variance as a function of input phase $\phi$ (unordered).
    (b) Same data after ordering.
    (c) Ordered and normalized variance exhibiting the expected $|\cos\phi|$ dependence with minima at quadrature.
    (d) Extracted residual phase error $\Delta\theta$ after calibration.}
    \label{fig:PS_calibration}
\end{figure}

Figure~\ref{fig:PS_calibration} summarizes the experimental calibration of a representative phase shifter (PS$_{63}$). Figures~\ref{fig:PS_calibration}(a) and (b) show the experimentally measured output variance as a function of $\phi$ before and after ordering the phase samples, respectively. The corresponding normalized variance is shown in Fig.~\ref{fig:PS_calibration}(c). The measured curve closely follows the theoretical $|\cos\phi|$ dependence predicted by Eq.~\eqref{eq:PS_sigma_main}, with well-defined variance minima observed near the quadrature points $\phi \approx \pi/2$ and $3\pi/2$. These minima identify the calibrated operating points of the phase shifter.

After fixing the heater voltage corresponding to the variance minimum, the residual phase error of the calibrated phase shifter is evaluated. Figure~\ref{fig:PS_calibration}(d) shows the extracted phase deviation $\Delta\theta$ as a function of the theoretical phase $\theta_{\mathrm{theory}}$ for this representative device. The residual phase error remains within approximately $\pm 0.1$~rad over the full tuning range, confirming reliable convergence of the variance-based PS calibration under experimental conditions. All phase shifters across the processor exhibit comparable behavior.

\subsection{System-Level Demonstration: Hadamard Transformation}

\begin{figure}[H]
    \centering
    \includegraphics[width=0.7\textwidth]{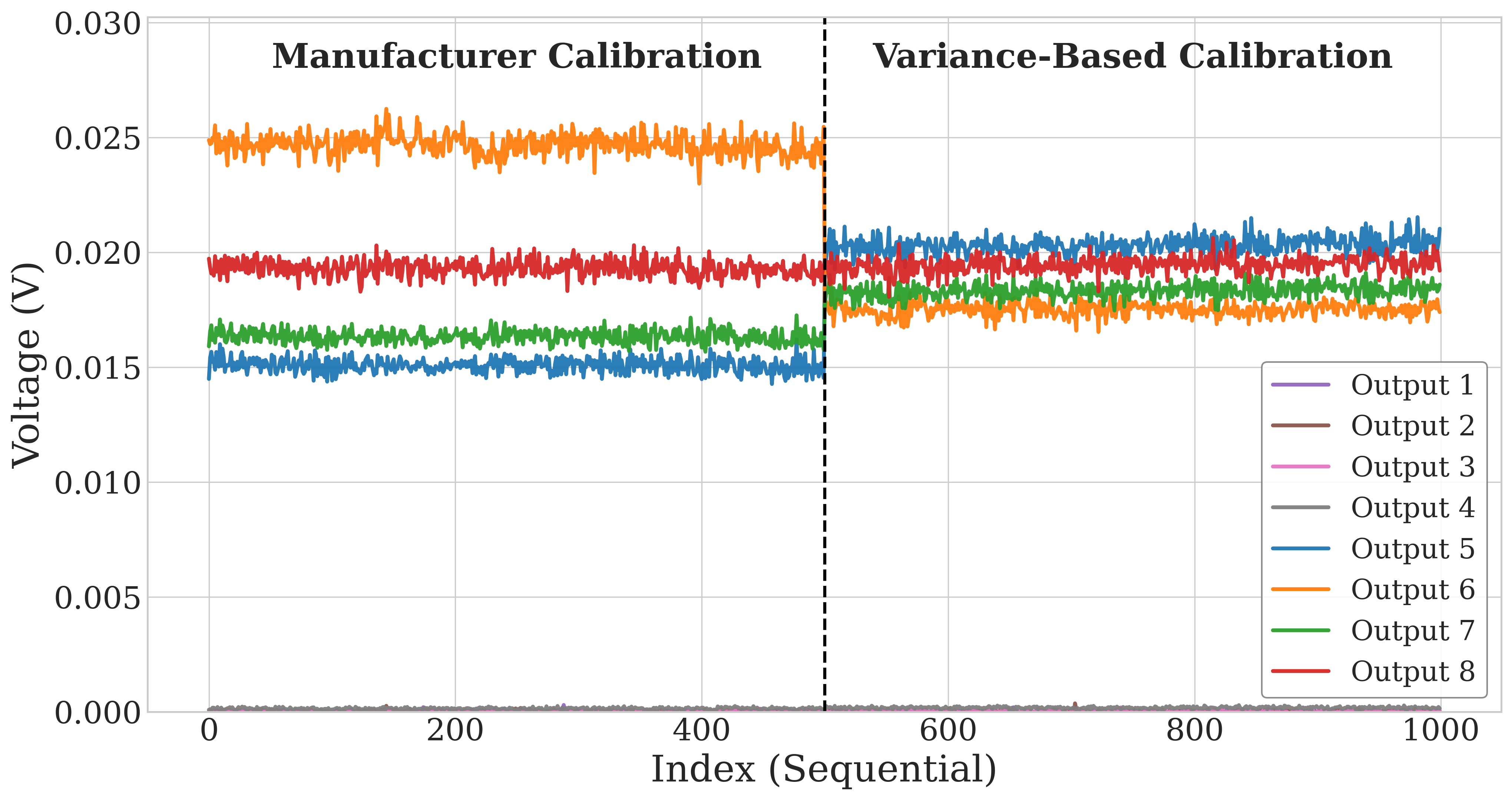} 
    \caption{System-level comparison between the manufacturer-provided factory
calibration and the proposed variance-based calibration for an implemented
$4\times4$ Hadamard transformation. The variance-based calibration yields
more uniform output levels and reduced temporal fluctuations across the
measured output channels.}
    \label{fig:power}
\end{figure}

To evaluate the impact of the proposed variance-based calibration method at the
system level, we implement an embedded $4\times4$ Hadamard transformation within
the fully calibrated $8\times8$ photonic processor using a Clements decomposition
of the target unitary. The remaining modes of the processor are left unused in
the logical transformation but are included in the physical implementation and
readout. The decomposition determines the required internal MZI splitting ratios and output phase settings, which are applied directly using the variance-based calibration parameters obtained in Sections~3.2 and 3.3, without applying any additional optimization or fine-tuning. To highlight the role of calibration, the same Hadamard configuration is also implemented using the manufacturer-provided factory calibration parameters, enabling a direct comparison between the default operating point and the variance-optimized operating point. For the experimental demonstration, coherent laser light at 1550~nm is injected into a single input port of the processor, while all other inputs are left unexcited, and the optical output voltages at all eight output ports are recorded sequentially using InGaAs photodiodes. Figure~\ref{fig:power} shows the recorded output signals under manufacturer calibration (left) and variance-based self-calibration (right). Under factory calibration, the output levels exhibit noticeable offsets and increased fluctuations, whereas the variance-based calibration yields significantly improved uniformity and temporal stability across all output channels. These results confirm that the proposed variance-based calibration method not only stabilizes individual interferometric building blocks but also translates directly into enhanced system-level performance for programmable photonic signal processing.

\section{Comparison with Existing Calibration Approaches}

In a programmable interferometric mesh, incorrect phase assignments not only disrupt the intended operation of individual unit cells but also propagate through the network and corrupt the optical states at all downstream elements. This sensitivity arises because the inputs to each unit cell are determined by the outputs of preceding units, rendering the global transfer function highly susceptible to calibration errors. For conventional sequential calibration to be accurate, two conditions must generally be satisfied: (i) only one input of the Mach--Zehnder interferometer (MZI) under characterization should be illuminated to suppress unwanted interference, and (ii) the neighboring interferometers must either have one dark input or be precisely biased into their bar or cross states. Enforcing these constraints typically requires explicit optical routing and careful control of the full mesh configuration.

Beyond node-isolation approaches, Miller~\cite{Miller2013} introduced a self-configuration framework based on the orthogonality of multiple simultaneously injected input fields. In this approach, each input beam corresponds to an orthogonal vector in the system Hilbert space, satisfying
\begin{equation}
\langle \mathbf{a}_i, \mathbf{a}_j \rangle = 0, \quad i \neq j ,
\end{equation}
so that tuning one channel does not perturb the others and large interferometric meshes can, in principle, be configured in parallel. While theoretically elegant, this orthogonality-based scheme requires precise global phase control and mutual coherence across all input ports. In practical integrated photonic circuits, such conditions are difficult to maintain due to fabrication-induced phase mismatches, thermal drift, and time-dependent phase noise. Moreover, generating and stabilizing orthogonal optical states introduces additional hardware complexity, including coherent splitters, phase-stabilized reference paths, and multi-channel synchronization.

A complementary multi-input strategy was proposed by Bandyopadhyay \textit{et al.}~\cite{Bandyopadhyay2021}, who introduced an averaging-based calibration method in which the phase of an upstream interferometer is swept over a full $2\pi$ range and the transmission of the interferometer under calibration is averaged over that sweep. This procedure suppresses the influence of upstream interference terms and yields an effective response proportional to $\sin^2(\theta/2)$, enabling accurate extraction of the internal phase. While this approach enables calibration in the presence of multiple simultaneously active inputs, it introduces a trade-off between accuracy and speed, since each MZI requires an additional $2\pi$ sweep of an upstream phase. As a consequence, the total calibration time scales linearly with the number of interferometers, and the method is sensitive to cumulative errors and thermal drift if the upstream elements are not already accurately calibrated.

Traditional node-isolation-based calibration techniques instead enforce specific optical paths through the processor to ensure that only the target interferometer is effectively excited. In rectangular meshes, this often involves diagonal routing from a selected input to a specific output, with neighboring interferometers biased into predefined states. While highly accurate, these procedures require repeated reconfiguration of the mesh, strict control of optical routing, and careful suppression of parasitic interference, all of which become increasingly complex as the processor size grows.

In contrast, the variance-based calibration method introduced here does not require conventional node isolation, orthogonal input modes, or auxiliary upstream phase sweeps. The calibration condition is inferred directly from the statistical variation of the measured output power during controlled phase perturbations. This removes the need for explicit optical routing, reference interferometers, or knowledge of the phase--voltage characteristics of the tuning elements. Calibration can be performed starting from fully random initial phase states and converges using only a small number of random phase realizations per tuning step.

In addition, the variance observable is separable across the
non-overlapping interferometers within a single vertical layer of the
Clements mesh. Provided that the corresponding output signals are
independently resolved and inter-element thermal crosstalk is negligible
or compensated, these interferometers can, in principle, be calibrated
concurrently. Unlike parallel calibration approaches that require
specifically calculated multiport input vectors, the locations of the
variance minima do not depend on knowledge of the local input amplitudes
or the absolute values of the applied phases.

Because the calibration observable is evaluated over an ensemble of
neighboring phase-shifter settings, the identified operating point is not
tied to a single neighboring configuration. Thermal offsets that remain
approximately constant over the ensemble and during the target-MZI scan
are incorporated into the measured variance minimum. The calibrated
voltage may therefore compensate for the average thermal load present
during calibration. In contrast, setting-dependent thermal crosstalk that
varies across the ensemble can introduce an additional variance floor and
may broaden or shift the measured minimum. Such effects require
experimental characterization and, where necessary, a subsequent
refinement step.

From a practical standpoint, this reduced control overhead directly translates
into simpler experimental implementations, faster calibration cycles, and
improved robustness to thermal drift and actuator nonlinearity.
While orthogonality-based and averaging-based methods remain powerful tools under highly controlled coherence conditions, the variance-based approach is particularly well suited for large-scale, thermally tuned, and dynamically reconfigurable photonic processors operating under realistic experimental constraints. In systems with strong global coherence, the different strategies may also be combined, using existing methods for coarse initialization followed by variance-based fine tuning for stable long-term operation.
\section{Conclusion}

We have introduced and experimentally demonstrated a variance-based self-calibration method for programmable photonic interferometer meshes that operates without node isolation, reference paths, orthogonal input states, or prior phase--voltage characterization. By exploiting the statistical properties of output intensity fluctuations under controlled phase perturbations, the proposed method enables direct identification of optimal bias points for both Mach--Zehnder interferometers and phase shifters using only intensity measurements.

We developed analytical models that establish explicit relationships between the measured output variance and the internal phase settings of the tuning elements, and we verified these models experimentally on an 8$\times$8 silicon-nitride photonic processor. The method was shown to converge reliably using only a small number of random phase samples, demonstrating strong robustness against experimental noise, thermal drift, and actuator nonlinearity. A two-stage automated calibration strategy was implemented in which all MZIs were first calibrated starting from fully random initial conditions, followed by variance-based phase-shifter calibration using balanced optical interference.

At the system level, the effectiveness of the proposed approach was validated through the implementation of an embedded $4\times4$ Hadamard transformation on the
$8\times8$ processor using a Clements decomposition. Compared to the manufacturer-provided factory calibration, the variance-based self-calibration yielded substantially improved output stability and uniformity across all processor output channels, confirming that the benefits of the method propagate from individual interferometric building blocks to full-scale programmable signal processing.

In comparison with existing calibration strategies based on node isolation, orthogonal input modes, or temporal averaging, the variance-based approach offers a unique combination of experimental simplicity, hardware efficiency, and robustness to phase instability. It is particularly well suited for large-scale, thermally tuned, and dynamically reconfigurable photonic processors operating under realistic experimental conditions.

These results establish variance-based calibration as a powerful and scalable alternative for the stable operation of programmable photonic hardware. Beyond classical linear optical signal processing, the method is directly applicable to emerging photonic computing, neuromorphic, and quantum photonic platforms, where robust, low-overhead, and automated calibration is essential for practical deployment.

\section*{Acknowledgment}
This work was supported by the Federal Ministry of Research,
Technology and Space of Germany through the Q-TREX project,
grant number 16KISR026, and the QD-CamNetz project, grant number
16KISQ077. This research is part of the Munich Quantum Valley,
which is supported by the Bavarian state government with funds
from the Hightech Agenda Bayern Plus.


\clearpage

\setcounter{section}{0}
\renewcommand{\thesection}{S\arabic{section}}

\setcounter{subsection}{0}
\renewcommand{\thesubsection}{S\arabic{section}.\arabic{subsection}}

\setcounter{equation}{0}
\renewcommand{\theequation}{S\arabic{equation}}

\setcounter{figure}{0}
\renewcommand{\thefigure}{S\arabic{figure}}

\setcounter{table}{0}
\renewcommand{\thetable}{S\arabic{table}}

\begin{center}
    {\LARGE\bfseries Supplementary Material}

    \vspace{0.4cm}

    {\large
    Local Variance-Based Calibration of Programmable Photonic
    Interferometer Meshes
    }

    \vspace{0.25cm}

    G{\"o}khan Elmas, Igor A. Litvin, and Janis N{\"o}tzel
\end{center}

\vspace{0.6cm}

\section{Variance-Based Mach--Zehnder Interferometer Calibration}
\label{supp:mzi_derivation}
Each unit cell in the photonic processor consists of a tunable Mach--Zehnder interferometer (MZI) followed by a phase shifter on one output arm. We denote the internal MZI phase as $\theta$ and the output-arm phase as $\varphi$. The scattering matrix of the tunable unit cell is
\begin{equation}
S(\theta,\varphi) = \frac{1}{2}
\begin{pmatrix}
1 - e^{-i\theta} & -i(1+e^{-i\theta}) \\
-i(1+e^{-i\theta})e^{-i\varphi} & -(1 - e^{-i\theta})e^{-i\varphi}
\end{pmatrix}.
\end{equation}

\subsection*{Input--Output Relation}

We inject a two-port input field
\begin{equation}
\mathbf{a}(\phi) =
\begin{pmatrix}
A e^{i\phi} \\
B
\end{pmatrix}, \qquad A,B>0, \;\; \phi \in [0,2\pi),
\end{equation}
where $\phi$ is the controllable relative input phase. The output field is
\begin{equation}
\mathbf{b}(\theta,\varphi,\phi) = S(\theta,\varphi)\mathbf{a}(\phi).
\end{equation}

The first output component is
\begin{equation}
b_1(\theta,\phi) = S_{11}(\theta,\varphi)A e^{i\phi} + S_{12}(\theta,\varphi)B,
\end{equation}
where
\begin{equation}
S_{11}(\theta,\varphi)=\frac{1-e^{-i\theta}}{2}, \qquad
S_{12}(\theta,\varphi)=-\frac{i(1+e^{-i\theta})}{2}.
\end{equation}
Note that the output phase $\varphi$ does not affect $b_1$.

\subsection*{Output Intensity}

The output intensity at port~1 is
\begin{align}
I(\theta,\phi)
&= |b_1(\theta,\phi)|^2 \\
&= \left|\frac{1-e^{-i\theta}}{2}A e^{i\phi}
- \frac{i(1+e^{-i\theta})}{2}B\right|^2.
\end{align}
Expanding explicitly,
\begin{equation}
I(\theta,\phi)
= \frac{A^2+B^2}{2}
+ \frac{B^2-A^2}{2}\cos\theta
- AB\sin\theta\cos\phi.
\end{equation}

\subsection*{Mean Over Input Phase}

Averaging over a uniform phase sweep $\phi\in[0,2\pi)$,
\begin{align}
\langle I(\theta,\cdot)\rangle_\phi
&= \frac{1}{2\pi}\int_0^{2\pi} I(\theta,\phi)d\phi \\
&= \frac{A^2+B^2}{2}+\frac{B^2-A^2}{2}\cos\theta,
\end{align}
since $\langle\cos\phi\rangle_\phi=0$.

\subsection*{Variance Over Input Phase}

The deviation from the mean is
\begin{equation}
I(\theta,\phi)-\langle I(\theta,\cdot)\rangle_\phi
= -AB\sin\theta\cos\phi.
\end{equation}
Squaring and averaging,
\begin{align}
\mathrm{Var}_\phi[I(\theta,\cdot)]
&= \frac{1}{2\pi}\int_0^{2\pi}
A^2B^2\sin^2\theta\cos^2\phi\,d\phi \\
&= \frac{A^2B^2\sin^2\theta}{2},
\end{align}
which yields
\begin{equation}
\sigma(\theta)=\frac{AB}{\sqrt{2}}|\sin\theta|.
\end{equation}

\subsection*{Finite Random Phase Sampling}

For a finite random phase set $\Phi_M=\{\phi_1,\dots,\phi_M\}$,
\begin{equation}
\mathrm{Var}_{\Phi_M}[I(\theta,\cdot)]
= A^2B^2\sin^2\theta\,C_M,
\end{equation}
with
\begin{equation}
C_M = \frac{1}{M}\sum_{k=1}^M\cos^2\phi_k
-\left(\frac{1}{M}\sum_{k=1}^M\cos\phi_k\right)^2.
\end{equation}
The finite-sample standard deviation becomes
\begin{equation}
\sigma^{(M)}(\theta)=AB|\sin\theta|\sqrt{C_M}.
\end{equation}
For independent uniformly distributed phase samples and the empirical variance
normalization used here, $\mathbb{E}[C_M]=(M-1)/(2M)$, which approaches $1/2$
as $M\to\infty$.

\section{Variance-Based Phase-Shifter Calibration}
\label{supp:ps_derivation}

The same unit cell is now considered, but the goal is to calibrate the
phase-shifter-controlled relative phase $\phi$, while the auxiliary MZI phase
$\theta$ is sampled. The scattering matrix is again
\begin{equation}
S(\theta,\varphi)=\frac{1}{2}
\begin{pmatrix}
1 - e^{-i\theta} & -i(1+e^{-i\theta}) \\
-i(1+e^{-i\theta})e^{-i\varphi} & -(1 - e^{-i\theta})e^{-i\varphi}
\end{pmatrix}.
\end{equation}

\subsection*{Input--Output Relation}

We inject
\begin{equation}
\mathbf{a}(\phi)=
\begin{pmatrix}
A e^{i\phi} \\
B
\end{pmatrix},
\qquad \phi\in[0,2\pi).
\end{equation}
The output is
\begin{equation}
\mathbf{b}=S(\theta,\varphi)\mathbf{a},
\end{equation}
and the first port field is
\begin{equation}
b_1(\theta,\phi)=S_{11}(\theta)A e^{i\phi}+S_{12}(\theta)B,
\end{equation}
which is independent of $\varphi$.

\subsection*{Output Power}

The output power is again
\begin{equation}
P_1(\theta,\phi)
= \frac{A^2+B^2}{2}
+ \frac{B^2-A^2}{2}\cos\theta
- AB\sin\theta\cos\phi.
\end{equation}

\subsection*{Mean Over Phase-Shifter Fluctuations}

Assuming $\theta$ is uniformly distributed,
\begin{equation}
\langle P_1\rangle_\theta=
\frac{1}{2\pi}\int_0^{2\pi}P_1(\theta,\phi)d\theta
= \frac{A^2+B^2}{2}.
\end{equation}

\subsection*{Variance Over Phase-Shifter Fluctuations}

The deviation is
\begin{equation}
P_1(\theta,\phi)-\langle P_1\rangle_\theta
= \frac{B^2-A^2}{2}\cos\theta
-AB\sin\theta\cos\phi.
\end{equation}
Squaring and averaging,
\begin{align}
\mathrm{Var}_\theta[P_1(\cdot,\phi)]
&= \frac{1}{2\pi}\int_0^{2\pi}
\left[\frac{B^2-A^2}{2}\cos\theta
-AB\sin\theta\cos\phi\right]^2 d\theta \\
&= \frac{(A^2-B^2)^2}{8}
+ \frac{A^2B^2}{2}\cos^2\phi.
\end{align}

For balanced inputs $A=B$, this simplifies to
\begin{equation}
\mathrm{Var}_\theta[P_1(\cdot,\phi)]
= \frac{A^4}{2}\cos^2\phi,
\end{equation}
giving
\begin{equation}
\sigma(\phi)=\frac{A^2}{\sqrt{2}}|\cos\phi|.
\end{equation}

\subsection*{Finite Random Phase Sampling}

With a finite random phase-shifter set $\Theta_M=\{\theta_1,\dots,\theta_M\}$, the same functional dependence on $\phi$ is recovered up to a constant scaling factor, yielding the normalized relation
\begin{equation}
\sigma_{\mathrm{norm}}(\phi)=|\cos\phi|.
\end{equation}
The variance therefore vanishes at $\phi=\pi/2$ and $3\pi/2$, identifying the quadrature operating points used for phase-shifter calibration.

\end{document}